\documentclass[preprint,groupedaddress, showkeys,showpacs]{revtex4}

\usepackage{graphicx}
\usepackage{amsmath}
\usepackage{mathrsfs}

\begin{document}


\title{Chirality separation of mixed chiral microswimmers in a periodic channel}
\author{Bao-quan  Ai $^{1}$} \email[Email: ]{aibq@scnu.edu.cn}
\author{Ya-feng He$^{2}$}    \email[Email: ]{heyf@hbu.edu.cn}
 \author{Wei-rong Zhong$^{3}$}  \email[Email: ]{wrzhong@jnu.edu.cn}
\affiliation{$^{1}$Guangdong Provincial Key Laboratory of Quantum Engineering and Quantum Materials, School of Physics and Telecommunication
Engineering, South China Normal University, 510006 Guangzhou, China\\
$^{2}$College of Physics Science and Technology, Hebei University, 071002 Baoding, China.\\
$^{3}$Department of Physics and Siyuan Laboratory, College of Science and Engineering, Jinan University, 510632 Guangzhou, China.}


\date{\today}
\begin{abstract}
  \indent Dynamics and separation of mixed chiral microswimmers are numerically investigated in a channel with regular arrays of rigid half-circle obstacles. For zero shear flow, transport behaviors are the same for different chiral particles: the average velocity decreases with increase of the rotational diffusion coefficient, the direction of the transport can be reversed by tuning the angular velocity, and there exists an optimal value of the packing fraction at which the average velocity takes its maximal value. However, when the shear flow is considered, different chiral particles show different behaviors. By suitably tailoring parameters, particles with different chiralities can move in different directions and be separated. In addition, we also proposed a space separation method by introducing a constant load, where counterclockwise and clockwise particles stay in different regions of the channel.
  \end{abstract}

\pacs{82. 70. Dd, 05. 40. -a, 05. 60. Cd}



\maketitle
\section {Introduction}
\indent  Active matter in biological and physical systems has been studied theoretically and experimentally\cite{rmp,rmp1,rpp}. Active particles or agents are assumed to have an internal propulsion mechanism, which may use energy from an external source and transform it under non-equilibrium conditions into directed motion. Understanding this kind of motion can provide insight into out-of-equilibrium phenomena associated with biological examples such as bacteria, as well as with artificial microswimmers. These microswimmers may appear in several fields\cite{rmp1,rpp}, e.g., to perform tasks in lab-on-a-chip devices, to localize pollutants in soils, and to deliver drugs within tissues. Compared with passive Brownian particles, active particles moving in confined structures could exhibit peculiar behaviors\cite{Leonardo,Wan,Ghosh,angelani,Ai,Buttinoni,Schwarz-Lineka,Fily,Kaiser,Rusconi,Wioland,Vicsek1,Nepusz,Volpe,Costanzo,Maggi,Yang,McCandlish,Berdakin,DiLuzio,Li,Ao,Mijalkov,Reichhardt,Nguyen,ohta,peruani,stark,Zhang,nc,Hagen,Potiguar}, resulting for example in spontaneous rectified transport\cite{Leonardo,Wan,Ghosh,angelani,Ai}, phase separation of self-propelled particles\cite{Buttinoni,Schwarz-Lineka,Fily}, trapping of particles in microwedge\cite{Kaiser}, depletion of elongated particles from low-shear regions\cite{Rusconi}, spiral vortex formation in circular confinement\cite{Wioland}, and collective motion in complex systems\cite{Vicsek1,Nepusz} .

\indent Separation strategies of active particles based on their swimming properties \cite{Volpe,Costanzo,Maggi,Yang}are of utmost importance for various branches of science and engineering. Costanzo and coworkers \cite{Costanzo} proposed a separation method for run-and-tumble particles in terms of their motility in a microchannel. The centrifugation can be also used to separate swimming cells having different motility\cite{Maggi}. Yang et. al.\cite{Yang} proposed to use self-driven artificial microswimmers for the separation of binary mixtures of colloids. McCandlish and coworkers\cite{McCandlish} studied the spontaneous segregation of active and passive particles, freely swimming in a two-dimensional box. Asymmetric obstacles
can also induce the separation of self-propelled particles with different motilities \cite{Berdakin}. These separation strategies were mainly based on different motilities. However, in some situations, chirality separation is very important, for example often only one specific chirality is needed by the pharmaceutical and chemical industry. Therefore, a deeper understanding of  chiral sorting mechanisms is of great fundamental and technological importance. Recently, Mijalkov and Volpe\cite{Mijalkov} demonstrated that chiral microswimmers can be sorted on the basis of their swimming properties by employing some simple static patterns in their environment. In asymmetrically patterned arrays, particles with the same radius but different chirality move in different directions\cite{Reichhardt}.

\indent In this paper, we numerically study the dynamics and separation of mixed chiral particles in a channel with regular arrays of rigid half-circle obstacles. Two chiral separation methods are proposed: (A)Movement direction separation: in the presence of the shear flow, the counterclockwise particles move to the right while the clockwise particles move to the left. (B)Space separation: in the presence of a constant load, the counterclockwise particles stay in the upper region, while the clockwise particles stay in the bottom region.

\section{Model and methods}
\indent In the presence of an additional torque, the microswimmer tends to execute circular obits called as a chiral microswimmer. For example,  when E. coli bacteria and spermatozoa undergo helicoidal motion near boundaries, which becomes two-dimensional chiral active Brownian motion \cite{DiLuzio}.
The kinetic of chiral active particles in confined structures could exhibit peculiar behavior. In a channel, chiral mciroswimmers drift autonomously along a narrow channel under more general asymmetry conditions\cite{Li} and the diffusion of chiral mciroswimmers depends on both the strength and the chirality of the self-propulsion mechanism\cite{Ao}.

\begin{figure}[htbp]
\vspace{0cm}
\begin{center}
\includegraphics[width=8cm]{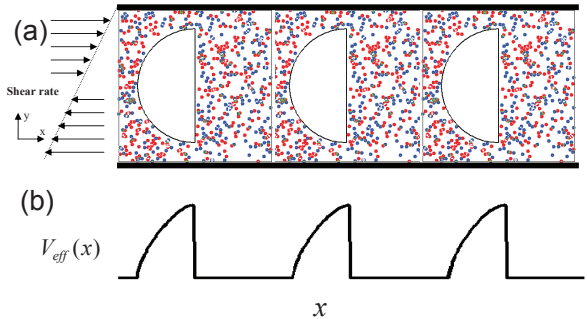}
\vspace{0cm}
\caption{(Color online) (a) Scheme of the separating device: two kinds of chiral self-propelled particles moving in Couette flow with shear rate $\gamma$. The half circle obstacles are regularly arrayed in the channel. Periodic boundary conditions are imposed in the $x$ direction, and hard wall boundaries in the $y$ direction. The red and blue balls denote counterclockwise and clockwise particles, respectively. (b)The effective entropic potential $V_{eff}(x)$: the reduction of the coordinates (i. e. elimination of $y$ coordinate by assuming equilibrium in the orthogonal direction) can involve the appearance of the entropic barrier.}\label{1}
\end{center}
\end{figure}
\indent In this paper, we consider mixed chiral particles moving in Couette flow\cite{Hagen} with walls (the height $H$) in the $y$-direction and periodic boundary conditions (the period $L$) in the $x$-direction shown in Fig. 1(a). Strictly, we do not consider microswimmers but active particles, because the hydrodynamic coupling between particle motion was ignored. The rigid half-circle obstacles with the diameter $D$ are regularly arrayed in the channel. The active particles are modeled as soft disks with the diameter $a$ and it is viewed as characterized by a unit vector $\vec{n}_i\equiv (\cos\theta_i,\sin\theta_i)$ in the $x-y$ plane, defining the direction of the self-propelled velocity. The effect of the shear flow enters the equations via the shear rate $\gamma$. The forces acting particle $i$ from the other particles and from the obstacles are respectively defined as $\mathbf{F}_i=F^{x}_{i}\mathbf{e}_x+F^{y}_{i}\mathbf{e}_y=\sum_{j}\mathbf{F}_{ij}$ and  $\mathbf{G}_{i}=G^{x}_{i}\mathbf{e}_x+G^{y}_{i}\mathbf{e}_y=\sum_{j}\mathbf{G}_{ij}$.
The dynamics of particle $i$ is described by the following overdamped Langevin equations
\begin{equation}\label{e1}
  \frac{dx_i}{dt}=\gamma y_i+v_0\cos\theta_i+\mu (F^{x}_{i}+G^{x}_{i}+f_{load}),
\end{equation}
\begin{equation}\label{e2}
  \frac{dy_i}{dt}=v_0\sin\theta_i+\mu (F^{y}_{i}+G^{y}_{i}),
\end{equation}

\begin{equation}\label{e3}
  \frac{d\theta_{i}}{dt}=-\frac{\gamma}{2}+\Omega_{i}+\sqrt{2D_{\theta}}\xi_{i}(t),
\end{equation}
where  $\mu$ is the mobility and $v_0$ is the magnitude of the self-propelled velocity.  $D_{\theta}$ is the rotational diffusion coefficient. $f_{load}$ is a constant load along $x$ direction. The Gaussian white noise $\xi_{i}(t)$ satisfies $\langle\xi_{i}(t)\rangle = 0$, and  $\langle \xi_{i}(t)\xi_{j}(s)\rangle = \delta_{ij}\delta(t-s)$. $\Omega$ is the angular velocity and the sign of $\Omega$ determines the chirality of the particle. We define particles as the counterclockwise particles for positive $\Omega$ and the clockwise particles for negative $\Omega$.

\indent The particle-particle interaction and particle-obstacle interaction are assume to be of the linear spring form with the stiffness constant $k_0$ and $k$, respectively. $\mathbf{F}_{ij}=k_0(a-r_{ij})\mathbf{e}_{r}$, if $r_{ij}<a$ ($\mathbf{F}_{ij}=0$ otherwise), and $r_{ij}$ is the distance between particle $i$ and $j$. $\mathbf{G}_{ij}=k[\frac{1}{2}(a+D)-r_{ij}]\mathbf{e}_{r}$, if $r_{ij}<\frac{1}{2}(a+D)$ ($\mathbf{G}_{ij}=0$ otherwise), and $r_{ij}$ is the distance between  particle $i$ and obstacle $j$( for a contact with the flat side of a half circle, $D=0$).

\indent Upon introducing characteristic length scale and time scale $\hat{x}=\frac{x}{a}$, $\hat{y}=\frac{y}{a}$, $\hat{t}=\mu k_0 t$, Eqs.(\ref{e1},\ref{e2},\ref{e3}) can be rewritten in the dimensionless forms
\begin{equation}\label{eq1}
  \frac{d\hat{x}_i}{d\hat{t}}=\hat{\gamma} \hat{y}_i+\hat{v}_0\cos\theta_i+\hat{F}_{i}^{\hat{x}}+\hat{G}_{i}^{\hat{x}}+\hat{f}_{load},
\end{equation}
\begin{equation}\label{eq2}
  \frac{d\hat{y}_i}{d\hat{t}}=\hat{v}_0\sin\theta_i+\hat{F}_{i}^{\hat{y}}+\hat{G}_{i}^{\hat{y}},
\end{equation}

\begin{equation}\label{eq3}
  \frac{d\theta_i}{d\hat{t}}=-\frac{\hat{\gamma}}{2}+\hat{\Omega_i}+\sqrt{2\hat{D}_{\theta}}\hat{\xi_i}(\hat{t}),
\end{equation}
and the other parameters can be rewritten as $\hat{\gamma}=\frac{\gamma}{\mu k_0}$, $\hat{v_0}=\frac{v_0}{\mu k_0a}$
$\hat{k}=\frac{k}{k_0}$, $\hat{\Omega}=\frac{\Omega}{\mu k_0}$, $\hat{D_{\theta}}=\frac{D_{\theta}}{\mu k_0}$, and $\hat{f}_{load}=\frac{f_{load}}{k_0a}$. From now on, we will use only the dimensionless variables and shall omit the hat for all quantities occurring in the above equations. Note that the model with zero shear flow and zero constant load was also studied in the previous work\cite{Potiguar}, where self-propelled particles can present a vortex-type motion around convex symmetric obstacles even in the absence of hydrodynamics effects.

\indent The ratio between the area occupied by particles and the total available area can be defined as the packing fraction $\phi=N\pi a^2/[4(LH-S_o)]$, where $S_o$ is the area covered by the obstacles and $N$ is the total number of particles. The system is simulated in a box with the above described boundary conditions. In the asymptotic long-time regime, the average velocity of particles along $x$ direction can be obtained from the following formula
           \begin{equation}\label{V}
            \bar{v}=\frac{1}{N}\sum_{i=1}^{N}\lim_{t\rightarrow\infty}\frac{[x_i(t)-x_i(0)]}{t}.
            \end{equation}
For convenience, we define the scaled average velocity $V_s=\overline{v}/v_0$ through the paper.

\indent In order to facilitate discussion, we can obtain the effective entropic potential after the elimination of $y$ coordinate by assuming equilibrium in the orthogonal direction and $V_{eff}(x)\propto-k_{B}\ln\frac{S(x)}{L}$, here $k_B$ is the Boltzmann constant and $S(x)$ is transverse cross section and defined as
 \begin{scriptsize}
 \begin{equation}\label{}
   S(x)=\left\{
\begin{aligned}
H &, & 0\leq x\leq \frac{1}{2}(L-D), \frac{L}{2}\leq x \leq L; \\
H-2\sqrt{(\frac{D}{2})^2-(x-\frac{L}{2})^2}&, & \frac{1}{2}(L-D)<x<\frac{L}{2},
\end{aligned}
\right.
 \end{equation}
 \end{scriptsize}
and the effective potential is shown in Fig. 1(b).

\section{Results and Discussion}
 \indent In this work, we propose two methods for chirality separation: (A) Movement direction separation by shear flow and (B)Space separation by a constant load. Unless otherwise noted, our simulations are under the parameter sets: $D=30$, $L=40$, $H=40$, and $k=100$. For the convenience of discussion, we focus on equimolar mixtures, where the number of the counterclockwise particles is equal to the number of the clockwise particles.

 \subsection{Chirality separation induced by shear flow}
\indent  In this section, we will systematically study the scaled average velocity $V_s$ in terms of $D_{\theta}$, $|\Omega|$, $\phi$, and $\gamma$ in the presence of shear flow at $f_{load}=0.0$.
\begin{figure}[htbp]
\begin{center}\includegraphics[width=8cm]{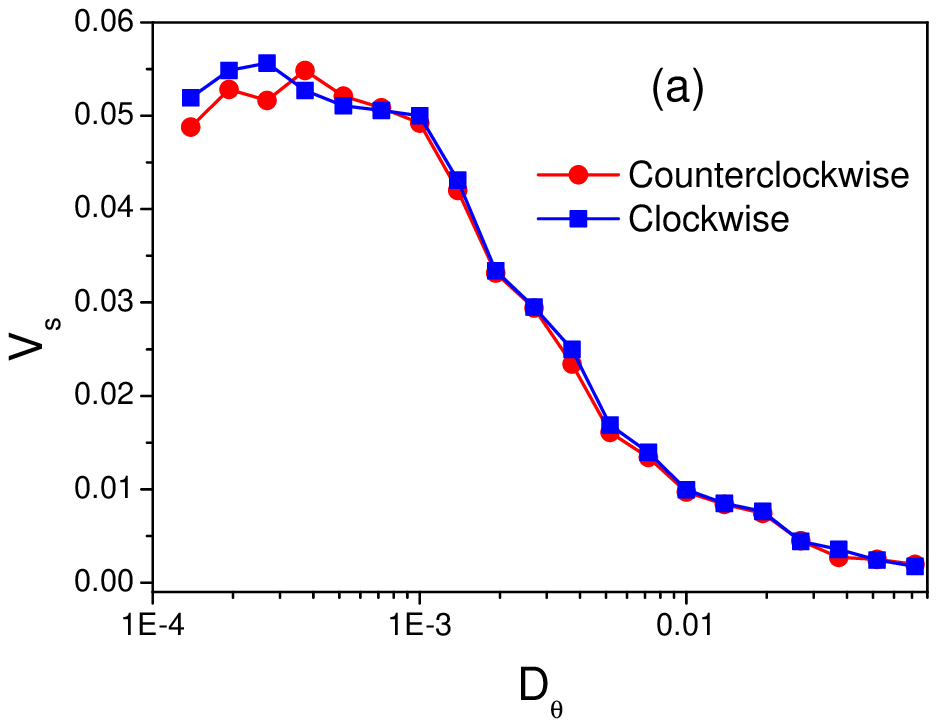}
\includegraphics[width=8cm]{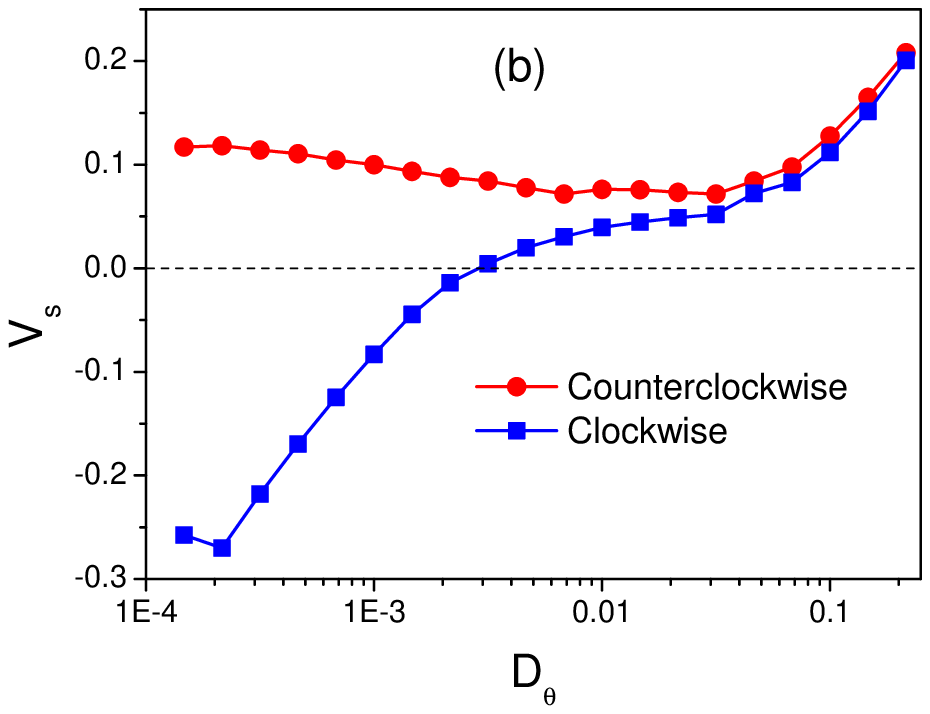}
\caption{(Color online) Average velocity $V_s$ as a function of the rotation diffusion $D_{\theta}$ at $\phi=0.1$. (a) $\gamma=0$ and $|\Omega|=0$. (b)$\gamma=0.004$ and $|\Omega|=0.02$. }\label{1}
\end{center}
\end{figure}

\indent Figure 2(a) shows the average velocity $V_s$ of nonchiral particles as a function of the rotation diffusion $D_{\theta}$  without the shear flow ($\gamma=0.0$). It is found that directed current occurs and the curves are almost the same for both clockwise (described by the blue line) and counterclockwise (described by the red line) particles. When $D_{\theta}$ is very small, the self-propelled angle $\theta$ almost does not change, $V_s$ tends to a saturate value. As $D_{\theta}$ increases, $V_s$ decreases monotonically. For very large $D_{\theta}$, particles change their directions very fast. The self-propelled velocity acts as a zero mean white noise and the average velocity then tends to zero. However, the transport behavior becomes different when the shear flow is considered (shown in Fig. 2(b)). On increasing $D_{\theta}$ from zero to 0.1, the average velocity of the clockwise particles increases monotonically from a negative value, while the average velocity of the counterclockwise particles decreases from a positive value. When $D_{\theta}>0.1$, the average velocities are the same for two kinds of particles. Obviously, particles in shear flow can be separated by the chirality only for small values of $D_{\theta}$(e. g. $D_{\theta}<0.002$). This behavior can be explained as follows. Due to the existence of the half circle, the counterclockwise and clockwise particles move easily in the up and down regimes of channel, respectively. When the shear flow is considered, the counterclockwise (clockwise) particles feel the positive (negative) shear force on average. Therefore, the counterclockwise particles move to the right while the clockwise particles move to the left.

\begin{figure}[htbp]
\begin{center}\includegraphics[width=8cm]{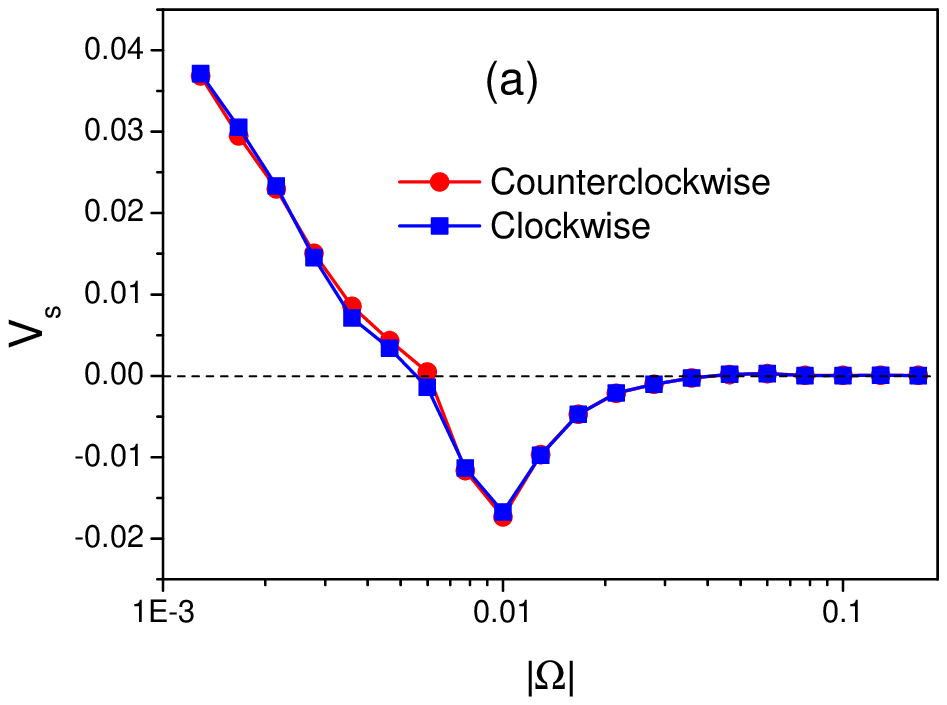}
\includegraphics[width=8cm]{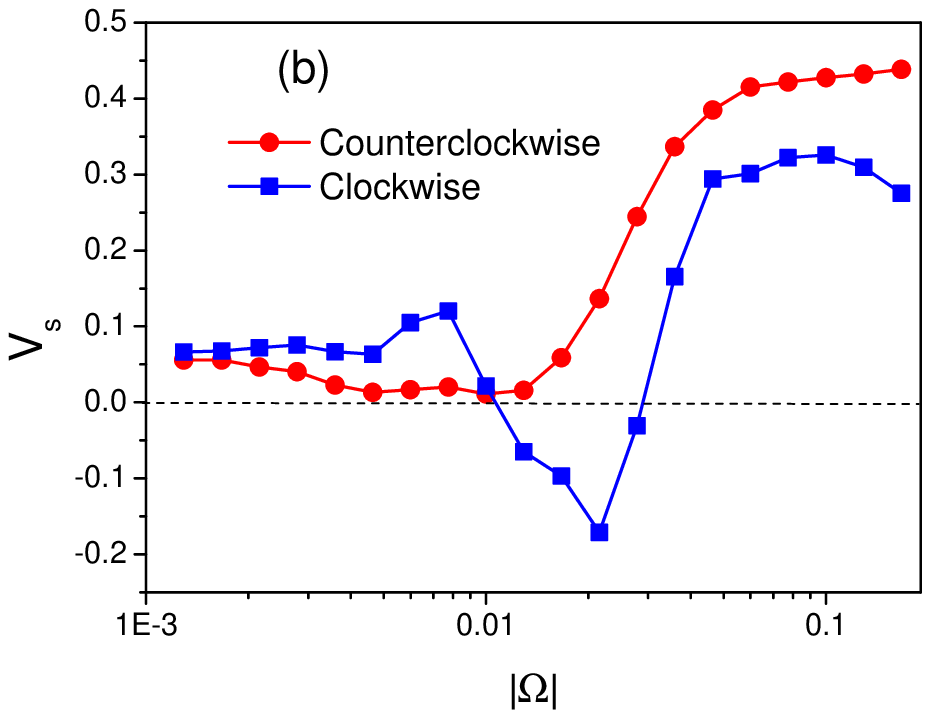}
\caption{(Color online)Average velocity $V_s$ as a function of the angular velocity $|\Omega|$ at $\phi=0.1$ and $D_{\theta}=0.0005$  . (a)$\gamma=0$. (b)$\gamma=0.004$.}\label{1}
\end{center}
\end{figure}

\indent The average velocity $V_s$ as a function of the angular velocity $|\Omega|$ is shown in Fig. 3. For zero shear flow ($\gamma=0.0$)(shown in Fig. 3(a)), the term $v_0 \cos \theta(t)$($ \propto v_0 \cos(|\Omega| t)$) in Eq. (\ref{e1}) can be seen as the external driving force and the effective potential is shown in Fig. 1(b).  In the adiabatic limit $|\Omega| \rightarrow 0$, the external force can be expressed by two opposite static forces $v_0$ and $-v_0$, yielding the mean velocity $V_s=\frac{1}{2}[V(v_0)+V(-v_0)]$. In this case, it is easier for particles moving toward the slanted side than toward the steeper side (see Fig.1(b)), so the average velocity is positive.  On increasing $|\Omega|$, due to the higher frequency, particles do not get enough time to cross the slanted entropic barrier (the right side) which is at a larger distance from the minima.  Since the distance from the minima to the basin of attraction of next minima from the steeper side (the left side) is less than from the slanted side (the right side), hence in one period particles get enough time to climb the entropic barrier from the steeper side than from the slanted side, which resulting in a negative current. When $|\Omega|\rightarrow \infty$, the self-propelled angle changes very fast, particles will experience a zero averaged constant force, so $V_s$ tends to zero. When the shear flow is involved, the average velocity of counterclockwise particles is always positive and it increases with the angular velocity. However, for the clockwise particles, the average velocity is positive for small and large values of $|\Omega|$, negative for medium values of $|\Omega|$. Therefore, there exists an interval of $|\Omega|$ where the counterclockwise particles move to the right while the clockwise particles move to the left.

\begin{figure}[htbp]
\begin{center}\includegraphics[width=8cm]{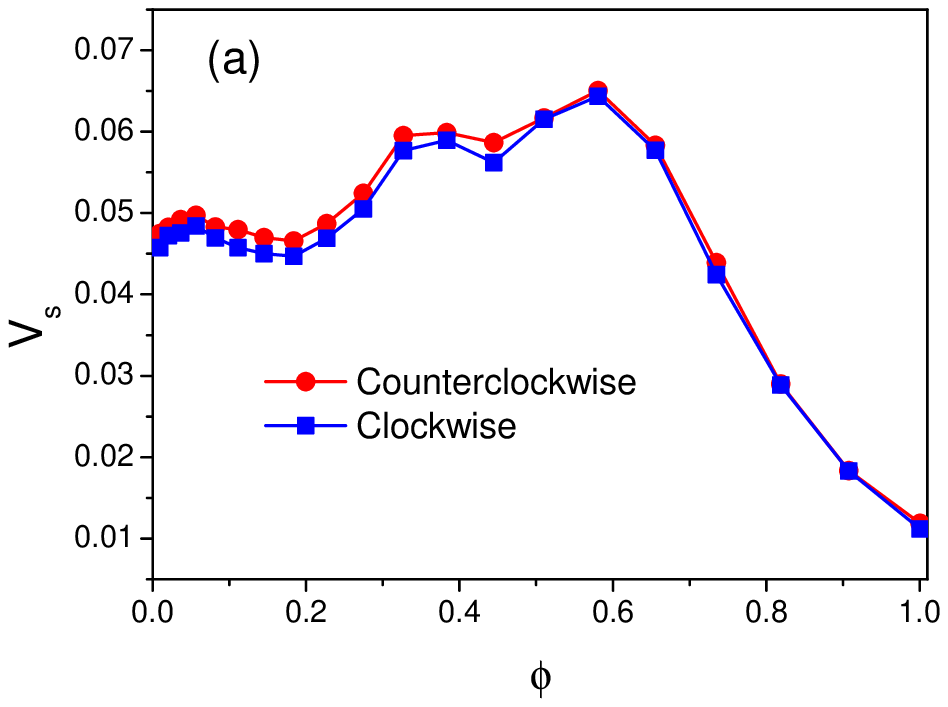}
\includegraphics[width=8cm]{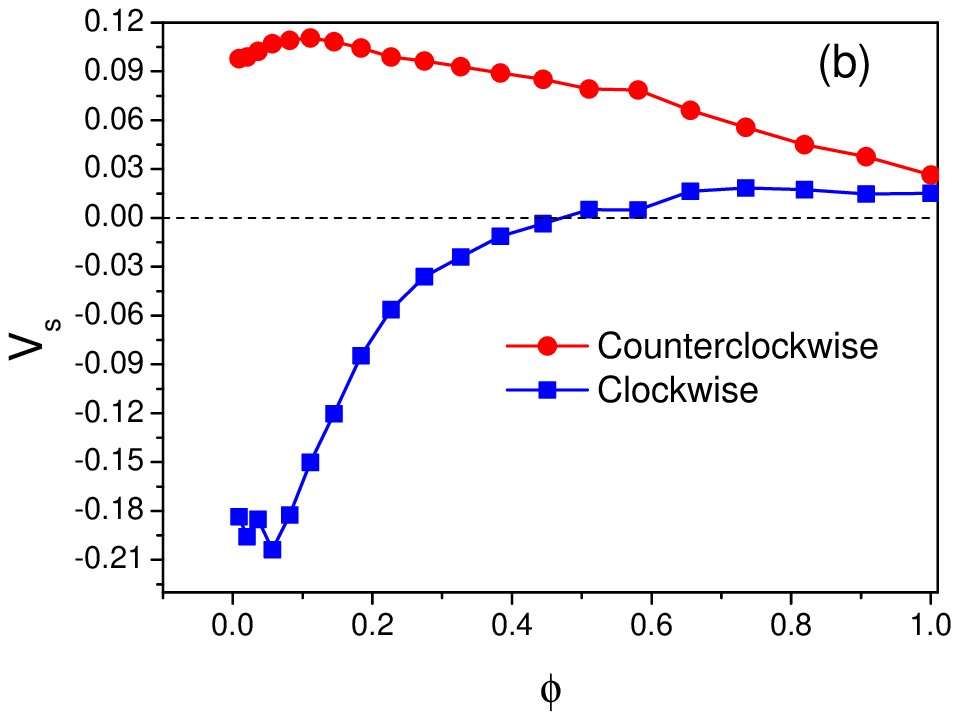}
\caption{(Color online) Average velocity $V_s$ as a function of the packing fraction $\phi$ at $D_{\theta}=0.0005$ . (a)$\gamma=0.0$ and $|\Omega|=0.0$. (b)$\gamma=0.004$ and $|\Omega|=0.02$.}\label{1}
\end{center}
\end{figure}

\indent Figure 4(a) shows the average velocity $V_s$ as a function of the packing fraction $\phi$ for zero shear flow.  The interaction between particles can cause two results: (A)activating motion in analogy with thermal noise activated motion for a single stochastically driven ratchet, which facilitates the ratchet transport and (B) reducing the self-propelled driving, which blocks the ratchet transport. When the packing fraction increases from zero, the factor A first dominates the transport, so the average velocity increases with the packing fraction. However, when $\phi>0.6$, the factor B becomes more important, the average velocity decreases with increasing $\phi$.  Therefore, there exists an optimal value of $\phi$ at which the average velocity takes its maximal value.  For the case of $\gamma=0.004$ and $|\Omega|=0.02$ (shown in Fig. 4(b)), we can find that when $\phi<0.4$ the counterclockwise particles move to the right while the clockwise particles move to the left (from Fig. 3(b)).  For this case, the average velocity of the counterclockwise particles is negative(positive) and its magnitude decreases (increases) with increasing $\phi$ for $\phi<0.4$($\phi>0.4$). On increasing $\phi$, the average velocity of the clockwise particles first increases, and then decreases.

\begin{figure}[htbp]
\begin{center}\includegraphics[width=7cm]{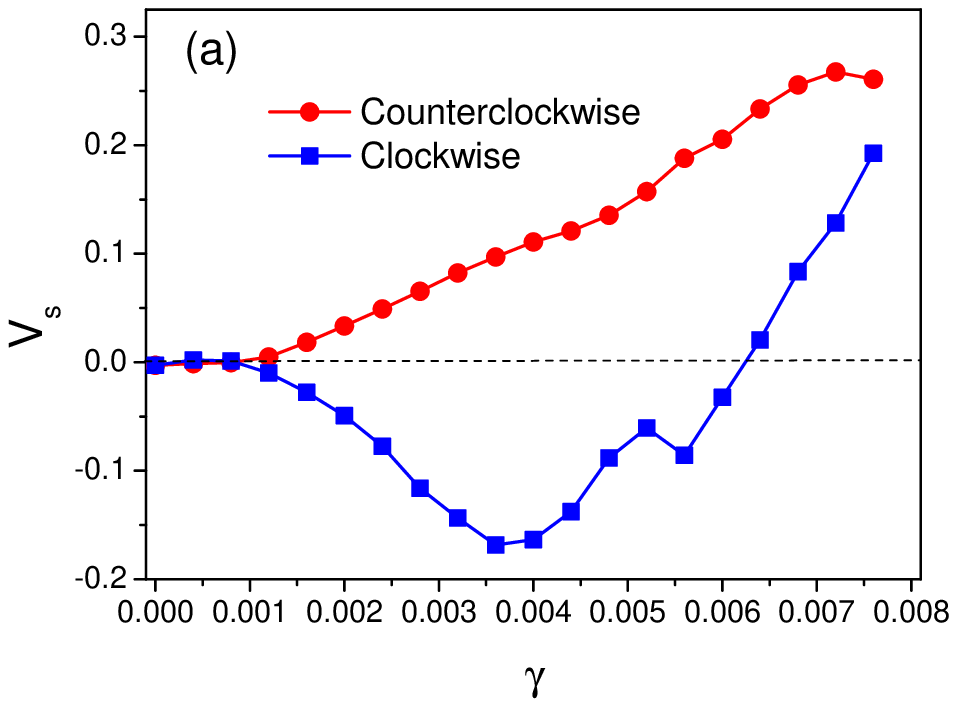}
\includegraphics[width=7cm]{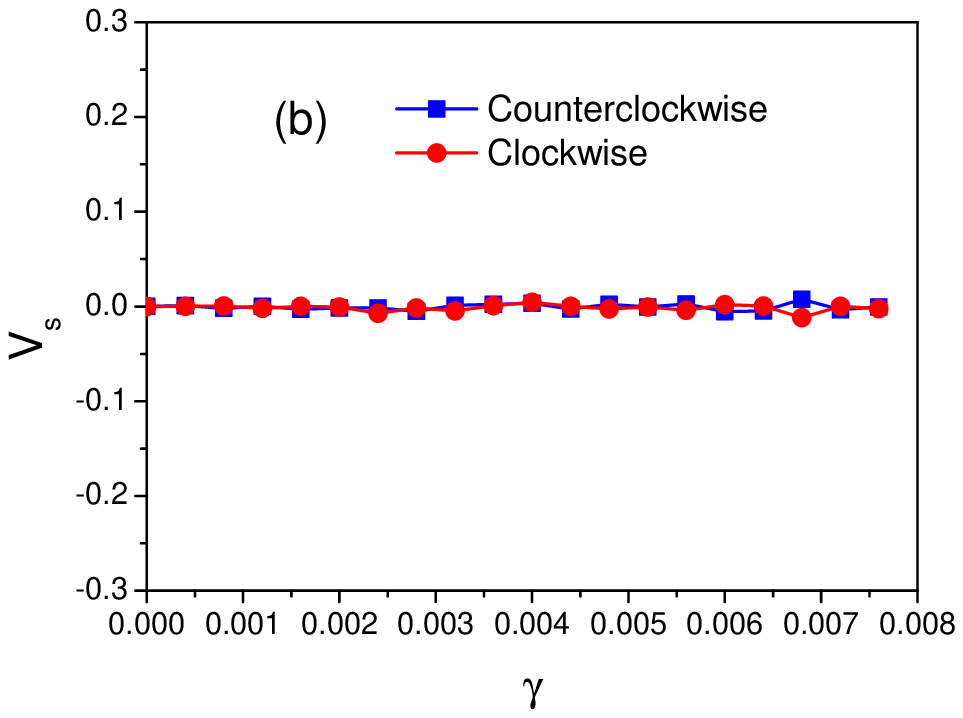}
\caption{(Color online)Average velocity $V_s$ as a function of the shear rate $\gamma$. (a)With half circle obstacles ($D=30.0$). (b)Without half circle obstacles ($D=0.0$). The other parameters are $|\Omega|=0.02$, $\phi=0.1$, and $D_{\theta}=0.0005$.}\label{1}
\end{center}
\end{figure}
\indent Figure 5(a) depicts the average velocity $V_s$ as a function of the shear rate $\gamma$ with half circle obstacles. It is found that the average velocity of the counterclockwise particles increases monotonously with the shear rate.  However, on increasing $\gamma$ from zero to $0.006$, the average velocity of the clockwise particles is negative and its magnitude first increases to its maxima, and then decreases to zero. When the shear rate is large enough(e. g. $\gamma>0.006$), the average velocities for both counterclockwise and clockwise particles are always positive. This is because the large shear rate dominates the transport and the chirality of the particle can be neglected. Therefore, chirality separation can be realized only for small shear rate.

\indent Note that both the shear flow and the obstacles are required for chirality separation in this setup. The shear flow alone does not suffice to drive the chiral species in opposite directions. For example, in the absence of the obstacles, the average velocity goes to zero for both counterclockwise and clockwise particles shown in Fig. 5(b). Obviously, without the obstacles, no net current occur even in the presence of the shear flow.

\indent Finally, we discuss the effect of the particle size on the transport when fixing the particle number($N=160$) shown in Fig. 6.  When $a\rightarrow 0$, the average velocity tends to a constant, which shows that the ratchet effect still appear even for point particles. When $a>(L-D)/2.0$, particles are too large to pass through the obstacles, thus the average velocity tends to zero. Therefore, there exists an optimal value of $a$ at which the average velocity takes its maximal value.
\begin{figure}[htbp]
\begin{center}\includegraphics[width=8cm]{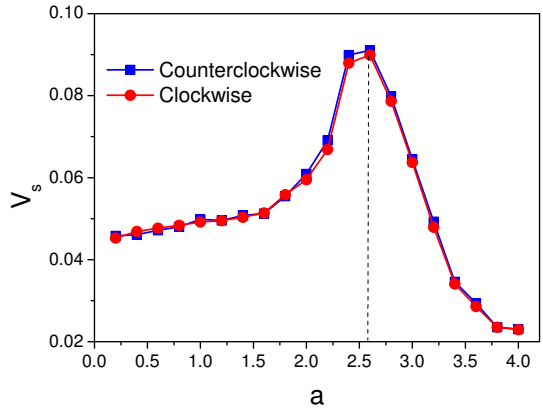}
\caption{(Color online)Average velocity $V_s$ as a function of the diameter $a$ of particles at $N=160$.  The other parameters are $|\Omega|=0.02$, $\gamma=0.0$, and $D_{\theta}=0.0005$.}\label{1}
\end{center}
\end{figure}

\subsection{Chirality separation driven induced by a constant load}
\indent In the above section, we proposed a movement direction separation method, where the counterclockwise particles move to the right while the clockwise particles move to the left under the special parameters. Now we will propose a space separation method where the counterclockwise and clockwise particles stay in different regions of the channel. In this method, we add a constant force $f_{load}$ along $x$ direction and no the shear flow is needed ($\gamma=0$).
\begin{figure}[htbp]
\begin{center}\includegraphics[width=7cm]{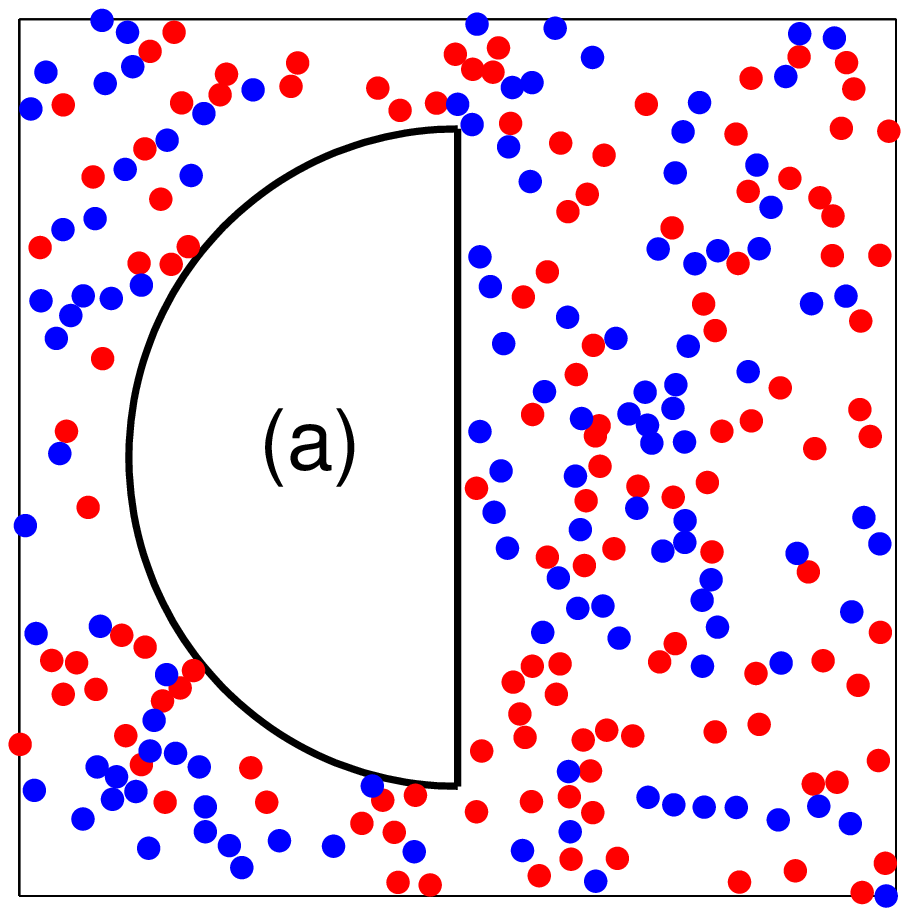}
\includegraphics[width=7cm]{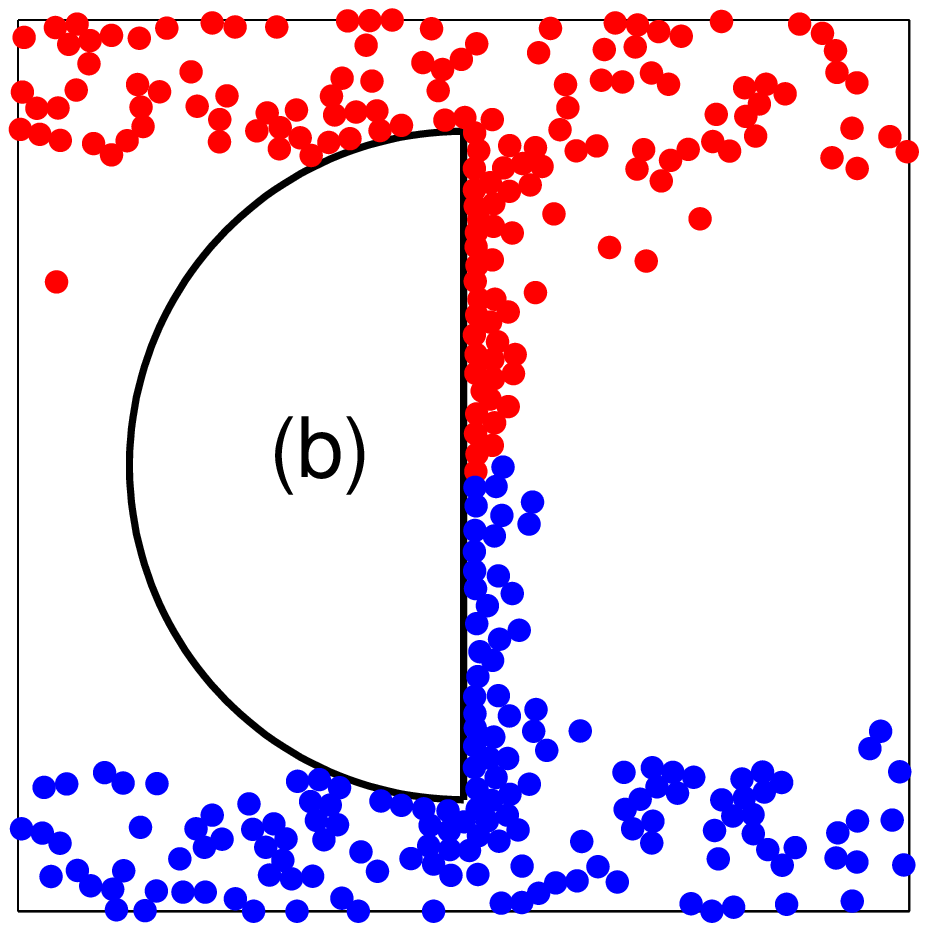}
\includegraphics[width=7cm]{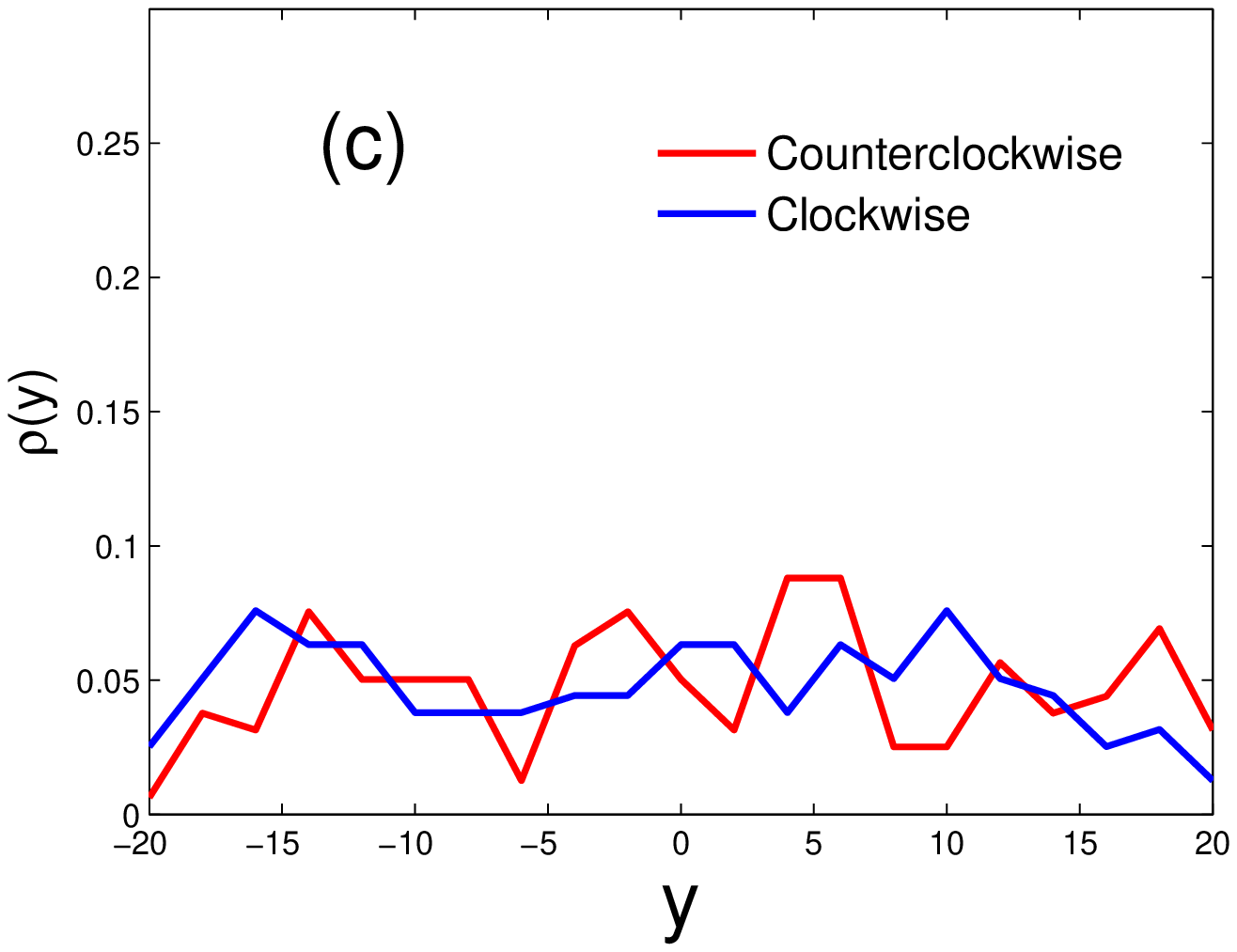}
\includegraphics[width=7cm]{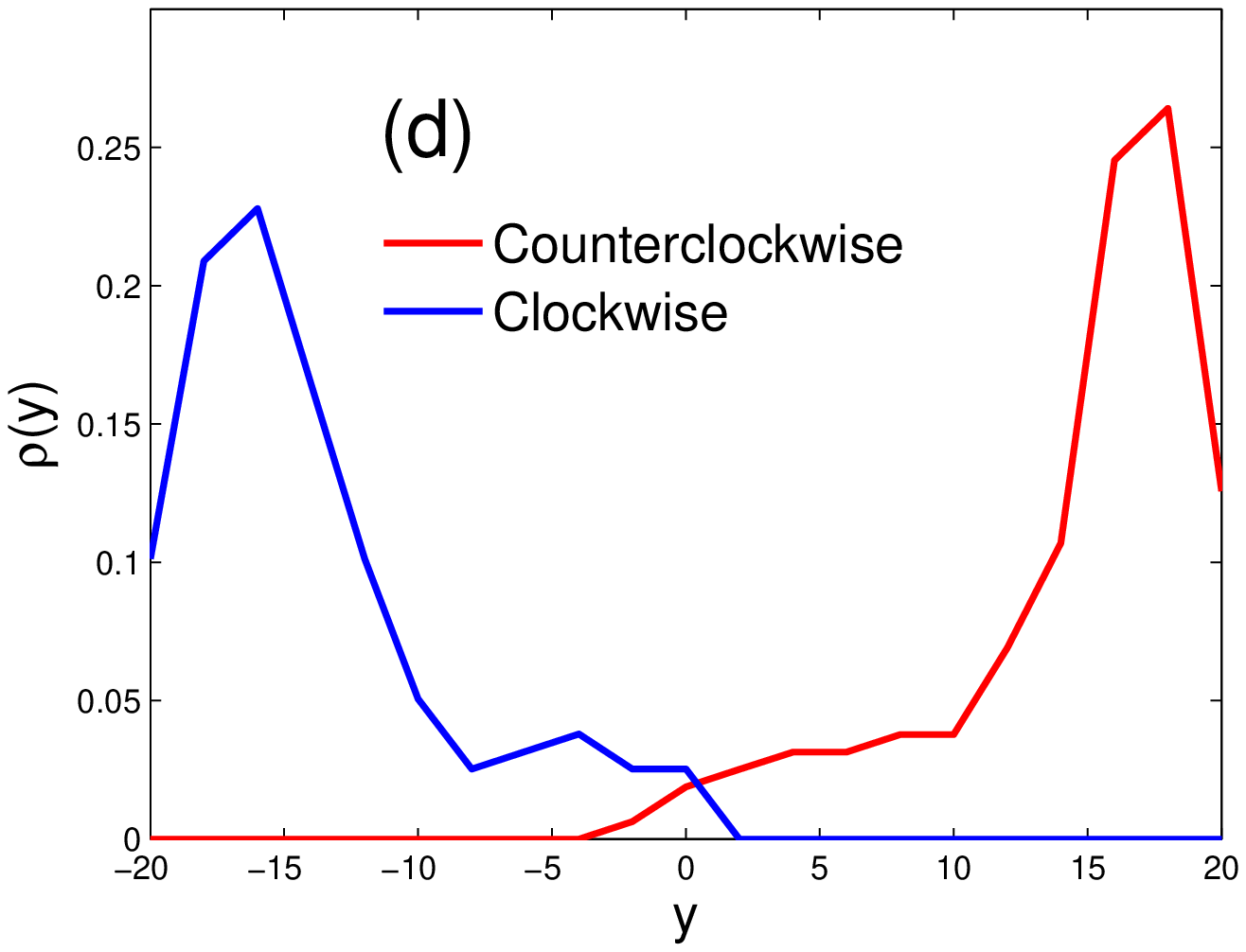}
\caption{(Color online)(a)Initial system configuration with uniform particle density at $t=0$. (b)The same system at $t=1000$. (c)Particle density $\rho(y)$ in $y$ direction at $t=0$. (d)Particle density $\rho(y)$ in $y$ direction at $t=1000$. The red (blue) circle denotes the counterclockwise (clockwise) particle. The other parameters are $|\Omega|=0.05$, $\phi=0.2$, $D_{\theta}=0.0005$, and $f_{load}=-0.1$.}\label{1}
\end{center}
\end{figure}

\indent Figure 7(a) shows the configuration of particles at $t=0$. The corresponding particle density in $y$ direction is shown in Fig. 7(c).  At $t=0$, both counterclockwise and clockwise particles are randomly distributed in the channel. These two kinds of particles are completely mixed together. After $t=1000$, the distribution of particles changes (shown in Figs. 7(b) and 7(d)): the counterclockwise particles are distributed in the upper region, while the clockwise particles stay in the bottom region.  Therefore, we can separation these two kinds of particles by space distribution. In order to give the explanation of the separation mechanism, we plot the schematic diagram of the space separation in Fig. 8.  As shown in Fig. 8, due to the existence of the load $f_{load}$, all particles move to the left on average. When particles meet the flat side of a half circle, different kinds of particles have different behaviors. The counterclockwise particles can easily pass across the half circle barrier through the upper region, while the clockwise particles pass across the barrier through the bottom region. Therefore, after a long time, the counterclockwise (clockwise) particles will be gathered in the upper (bottom) region.

\begin{figure}[htbp]
\vspace{0cm}
\begin{center}\includegraphics[width=4cm]{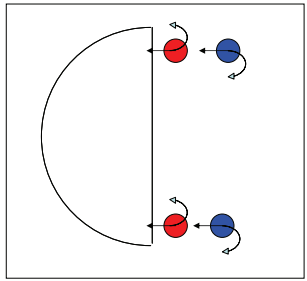}
\vspace{-0cm}
\caption{Schematic diagram of the space separation. The red (blue) circle denotes the counterclockwise (clockwise) particle.  }\label{1}
\end{center}
\end{figure}
\indent Note that the direction of the constant load is important for separating particles. The particle separation occurs only when the particles are driven towards the flat edges of the obstacles (e. g. $f<0$ in the present setup).  When particles are driven towards the rounded edges of the obstacles (e. g. $f>0$), two kinds of particles have almost the same space distribution (shown in Fig. 9), thus mixed particles can not be separated.

\begin{figure}[htbp]
\vspace{1cm}
\begin{center}
\includegraphics[width=7cm]{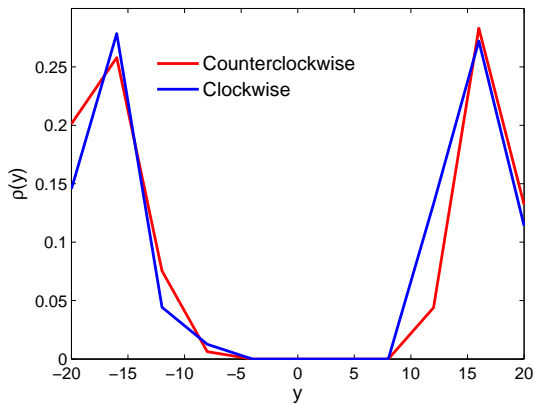}
\vspace{-0cm}
\caption{Particle density $\rho(y)$ in $y$ direction at $t=1000$. The other parameters are $|\Omega|=0.05$, $\phi=0.2$, $D_{\theta}=0.0005$, and $f_{load}=0.1$.  }\label{1}
\end{center}
\end{figure}

\subsection{Chirality separation in 3D systems}
\indent The separation mechanism for the case of two dimensional chiral particles performing circular motion can be extended to the case of three-dimensional (3D) chiral particles. The schematic diagram of the 3D setup is shown in Fig. 10, where mixed chiral particles moving in a channel with walls in the $y$ and $z$-directions and periodic boundary condition in the $x$-direction. The half-circle obstacle is replaced by the half-cylinder obstacle with the diameter $D$ and the height $L_z$ in 3D setup. The active particles are modeled as soft spheres with the diameter $a$ and the packing fraction is rewritten as $\phi=N\pi a^3/[6L_z(LH-0.125\pi D^2)]$. In the $x-y$ plane, we define particles as the counterclockwise particles for positive $\Omega$ and the clockwise particles for negative $\Omega$. To transfer the two-dimensional equations to the three dimensional case, the orientation vector $\vec{n}=(\sin\varphi_i\cos\theta_i, \sin\varphi_i\sin\theta_i, \cos\varphi_i)$ is used and the third component $z$ is added to all
quantities\cite{Hagen1}. The other parameters are the same as that in the two-dimensional case.

\begin{figure}[htbp]
\vspace{1cm}
\begin{center}
\includegraphics[width=7cm]{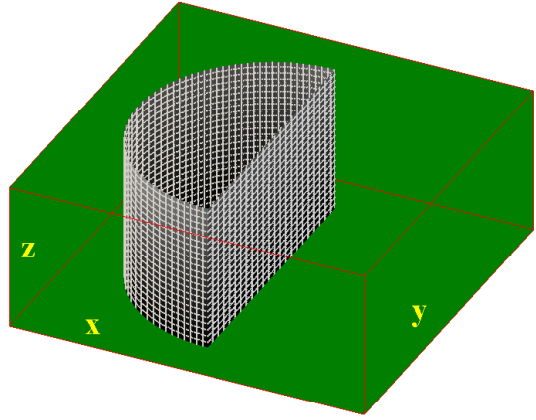}
\vspace{-0cm}
\caption{(Color online)Schematic diagram of the 3D separating device: two kinds of chiral self-propelled particles moving in the channel with walls in the $y$ and $z$-directions and periodic boundary condition in the $x$-direction. The half-cylinder obstacle is regularly arrayed in the channel. }\label{1}
\end{center}
\end{figure}

\begin{figure}[htbp]
\vspace{-0cm}
\begin{center}
\includegraphics[width=7cm]{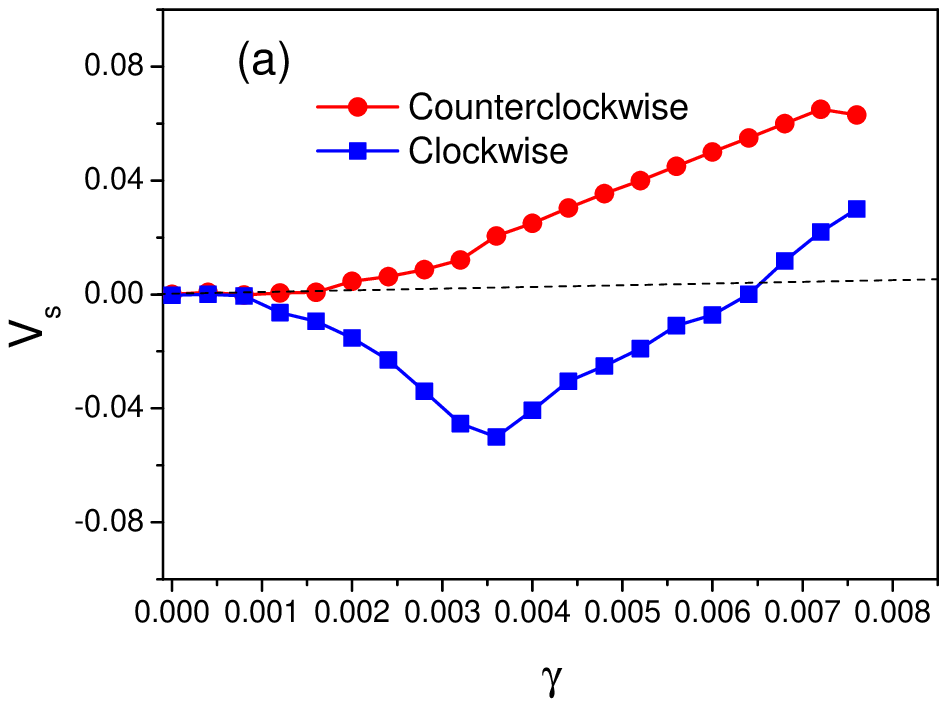}
\includegraphics[width=7cm]{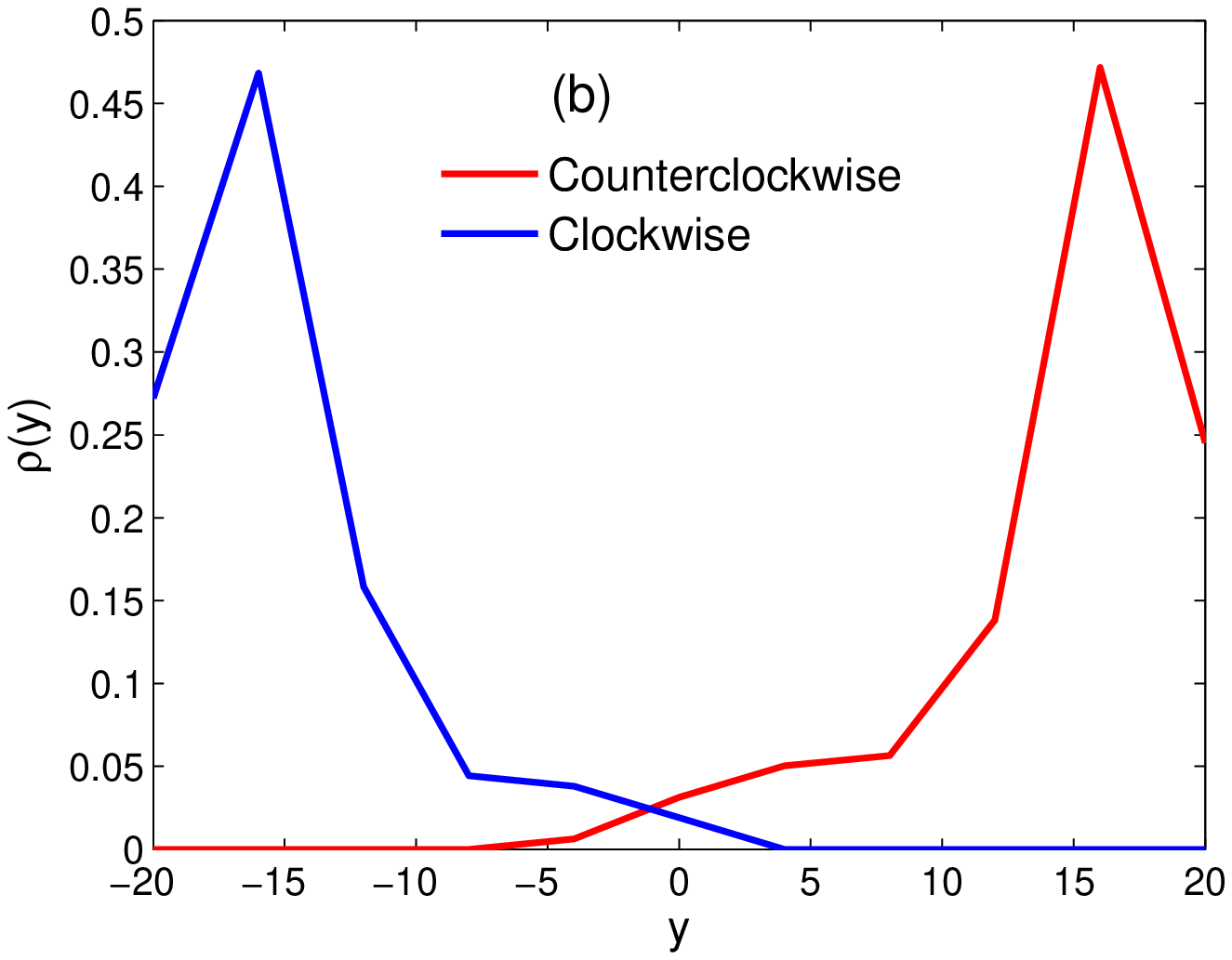}
\vspace{-0cm}
\caption{(Color online)Particle separation in the 3D setup. (a)Movement direction separation by the shear flow (corresponding to Fig. 5(a)). (b)Space separation by a constant load (corresponding to Fig. 7(d)). The red (blue) circle denotes the counterclockwise (clockwise) particle. The height of the half-cylinder $L_z=4.0$. The other parameters are the same as the corresponding 2D case.}\label{1}
\end{center}
\end{figure}
\indent The results for 3D separating devices are shown in Figs. 11(a) and 11(b).  It is found that two separation methods are still valid when $L_z<<L$ (e. g. $Lz=4.0$). When $L_z\rightarrow 0$, the 3D setup will reduce to 2D setup. From the extensive numerical simulations (not shown in the paper), the ratchet velocity $V_s$ decreases with increase of $L_z$. From the experimental point of view, the relatively large ratchet velocity $V_s$ is necessary for particle separation. Therefore, the condition of $L_z<<L$ must be considered when we design the 3D separating devices.

\indent From the above results, we can find that the optimal parameters for particle separation, $D_{\theta}<0.002$ and $0.01<|\Omega|<0.1$ in the dimensionless forms. From the experimental point of view, it would be interesting to estimate the real values of parameters. The characteristic parameters of Janus particles in a low Reynolds number regime\cite{Volpe} are $a=$2$\mu$m, $v_0=$10$\mu$m/s, and $\mu k_0=$100 s$^{-1}$.  From the scaling relations, one can obtain the real values of the rotational diffusion coefficient($D_{\theta}$$<$0.2 rad$^2$/s) and the angular velocity (1 rad/s$<|\Omega|<$10 rad/s). Obviously, it is possible to experimentally realize our separating mechanisms with these characteristic parameters.

\indent Note that our results are obtained from the projection of the half-cylinder in $x-y$ plane. However, in 3D situation,  a clockwise particle with its $z$ vector pointing upward is identical to a counterclockwise particle with its $z$ vector pointing down. Therefore, in order to improve the distinguishability of chiral swimming in 3D setup, a constant force $F_z$ is applied to the $z$ direction. In the presence of $F_z$, all particles accumulate at the upward (down) plane with $z=L_z$ ($z=0$) and can be easily distinguished.

\section {Concluding remarks}
\indent To conclude, we numerically studied the transport of mixed chiral particles moving in regular arrays of rigid half-circle obstacles.
The effect of the shear flow enters the equations via the shear rate $\gamma$. For zero shear flow, both counterclockwise and clockwise particles have the same transport behavior. The average velocity decreases with increase of the rotational diffusion coefficient. The direction of the transport can be reversed by tuning the angular velocity. There exists an optimal value of the packing fraction at which the average velocity takes its maximal value. However, when the shear flow is considered, transport behaviors for different chiral particles become different. By suitably tailoring the system parameters (the rotational diffusion $D_{\theta}$, the angular velocity $|\Omega|$, the packing fraction $\phi$, and the shear rate $\gamma$),  the counterclockwise particles move to the right while the clockwise particles move to the left. In addition, in order to obtain the space separation of chiral particles, we considered the constant load along the $x$-direction, instead of the shear flow. In this case, the counterclockwise particles are distributed in the upper region, while the clockwise particles stay in the bottom region, thus, we can separate chiral particles. The models are extended to the case of 3D chiral particles and the corresponding separating methods are still valid when $L_z<<L$. We expect that our separation methods can provide insight into out-of-equilibrium phenomena and have the technological applications for active microswimmers.

\indent  This work was supported in part by the National Natural Science Foundation of China (Grant Nos. 11175067, 11004082, and 11205044), the PCSIRT (Grant No. IRT1243),  the Natural Science Foundation of Guangdong Province(Grant No. 2014A030313426),  the program for Excellent Talents at the University of Guangdong Province, and the Fundamental Research Funds for the Central Universities, JNU (Grant Nos. 21611437 and 11614341).

\end{document}